\documentclass[
,12pt]{article}
\usepackage[affil-it]{authblk}
\usepackage[english]{babel}
\usepackage{blindtext}
\usepackage{enumitem}
\usepackage{indentfirst}

\providecommand{\keywords}[1]{\textbf{\textit{Keywords:}} #1}

\usepackage{graphicx}
\usepackage{amsmath}
\usepackage{amsthm}
\usepackage{amssymb}
\usepackage{amsfonts}
\usepackage[hidelinks]{hyperref}

\usepackage{booktabs}

\usepackage{longtable}
\usepackage{cite}
\usepackage{mathrsfs}

\usepackage[title,titletoc,toc]{appendix}

\newtheorem{case}{Case}
\newtheorem{subcase}{Subcase}

\theoremstyle{definition} 

\usepackage{chngcntr}
\counterwithin{subcase}{case}
\counterwithin{exmp}{section}
\counterwithin{rem}{section}
\counterwithin{red}{section}
\counterwithin{prof}{section}

\usepackage[margin=1in,left=1in,right=1in]{geometry}
\usepackage{subcaption}
\usepackage{float}
\usepackage{placeins}
\allowdisplaybreaks
\usepackage[final]{changes}
\usepackage{appendix}
\usepackage{cleveref}
\usepackage{longtable}
\usepackage{rotating}
\usepackage{array}
\usepackage{booktabs}

\title{Lie symmetry classification of a coupled nonlinear cross-diffusion system in radial geometry}

\author[1]{Manjit Singh\thanks{corresponding author: manjitcsir@gmail.com}}
\author[2]{Radhika}

\affil[1,2]{%
    Yadavindra Department of Sciences, Punjabi University Guru Kashi Campus, Talwandi Sabo--151302, Punjab, India.}
\begin{document}
\maketitle
\begin{abstract}
In this work, Lie symmetry analysis is performed on a coupled nonlinear cross-diffusion system with varying cross-section geometry. The system describes two interacting quantities whose material properties, namely the capacity functions and the diffusion coefficients, depend nonlinearly on the dependent variables. The classical Lie invariance criterion produces a set of sixteen determining equations for infinitesimal symmetry generators. The determining equations are solved by first establishing the universal geometric structure of the admitted generators and then classifying the constitutive functions according to their invariance properties in the state space. It is shown that the system always admits time translation and parabolic scaling as kernel symmetries, with an additional spatial translation admitted only in the Cartesian case. Further symmetries, such as translations, scalings, and rotations in the dependent-variable plane, are obtained by making precise structural assumptions about the constitutive functions. The analysis shows that the strong nonlinear coupling in the governing equations prohibits any new point symmetries from arising in the general case, and that larger symmetry algebras are only attainable in degenerate or linearizable special cases. The symmetries obtained in this work are geometrically consistent with parabolic and radial structure of governing  equations.

\end{abstract}

\section{Introduction}
In mathematical modelling of heat transfer processes, nonlinear heat equations with variable thermal coefficients emerge when the medium's thermal characteristics rely on state variables like temperature, concentration, density, or spatial position.
In contrast to the standard linear heat equation, which assumes constant thermal conductivity and diffusivity the nonlinear models include physically realistic features like temperature-dependent conductivity, anisotropic diffusion, phase transitions, and coupled transport phenomena.
From the perspective symmetry-analysis, variable thermal coefficients play a crucial role in determining the admitted Lie symmetries of the system. The presence of arbitrary functions generally reduce the admitted symmetries to only translation and scaling. However, special functional forms of the coefficients can expand the symmetry algebra, and hence more symmetry reductions, invariant solutions and conservation laws. As a result, classification of admissible functional coefficients is an essential topic in Lie group analysis of nonlinear heat equations.

The Lie group method is a powerful tool to investigate nonlinear differential equations for Lie symmetries. The theoretical foundation of Lie symmetry methods is well established in classical works such as Ovsiannikov \cite{ovsi}, Olver \cite{olverbook}, Bluman \cite{bluman2010applications}, Hydon \cite{haydonbook} and for algebraic structure of Lie algebras \cite{humphreys1972introduction,patera1977subalgebras}. The work of these authors outline a methodology for identifying admitted symmetry groups, constructing similarity variables, and converting partial differential equations to ordinary differential equations.

Lie symmetry analysis has been widely applied to heat-type and nonlinear evolution equations arising in mathematical physics. The research on nonlinear heat equation $u_{t}=(f(u)\,u_{x})_{x}$ was first started by Ovsiannikov in his seminal paper \cite{ovsiannikov1959group}. I.Sh. Akhatov, R.K. Gazizov and N.Kh. Ibragimov in \cite{akhatov1987group}, performed classification of $u_{t}=G(u_{x})u_{xx}$. V.A. Dorodnitsyn \cite{dorodnitsyn1982invariant} performed classification of $u_{t}=(G(u)u_{x})_{x}+g(u)$. Most extensive symmetry classification of nonlinear heat equation $u_{t}=(G(u)u_{x})_{x}+f(u)u_{x}$ was performed by A. Oron and P. Rosenau \cite{oron1986some} and M. P. Edwards, Maureen \cite{edwards1994classical} followed by generalization of nonlinear heat equation with convection term \cite{cherniha1998symmetries}. The group classification of most general class of heat equation $u_{t}=F(t,x,u,u_{x})\,u_{xx}+G(t,x,u,u_{x})$ is performed in \cite{zhdanov1999group,basarab2001structure,zhdanov2007group,abramenko2002group}. In \cite{yung1994group}, C.M. Yung, K. Verburg,P. Baveye performed group classification and symmetry reductions of the non-linear diffusion-convection equation $u_{t}=\,(D(u)\,u_{x})_{x}-K'(u)\,u_{x}$ and complete functional classification is performed. In \cite{cimpoiasu2008lie} the nonlinear heat equation $g(x)\,u_{t}=(f(u)u_{x})_{x}+h(u)\,u_{x}$ is investigated on similar framework for Lie symmetries corresponding to functional classes of $f(u)$ and $h(u)$. Recently, the study \cite{sinkala2025revisiting}  revisits and methodically re-derives the classical group classification of the nonlinear heat equation $u_{t}=({K}(u)\, u_{x})_{x}$, offering more precise and comprehensive derivations.

Nonlinear boundary-value and free-boundary problems that arise in mathematical physics and applied disciplines are another significant area of progress in Lie symmetry analysis. In monogram \cite{kumeibook}, Bluman and Kumei have given rigorous framework to apply symmetry methods to differential equations together with invariant boundary conditions and similarity reductions. Furthermore, Clarkson and Mansfield showed how symmetry approaches can be successfully applied to nonlinear evolution equations and associated boundary-value problems in \cite{clarkson1994algorithms}. These studies demonstrate that the compatibility of the admitted symmetry transformations with the related boundary and interface conditions is just as important to the efficacy of symmetry methods in applications as the invariance properties of the governing equations themselves.

The cross-diffusion system in cartesian such as following:
\begin{equation}
    \begin{aligned}
    C_{1}\,U_{t}=\,\frac{\partial}{\partial z}\bigg(K(u,v)\,u_{z}+L(u,v)\,v_{z}\bigg),\\
    C_{2}\,U_{t}=\,\frac{\partial}{\partial z}\bigg(M(u,v)\,u_{z}+D(u,v)\,v_{z}\bigg)
\end{aligned}
\end{equation}
represent diffusion across flat geometry, is widely studied in literature. This model is suitable for transport between parallel plates. However, many actual physical and biological processes do not occur in flat geometry. Most of the physical diffusion occurs where geometry has curvature, such as: heat conduction through a cylindrical pipe wall, groundwater flow toward a cylindrical or spherical well, heat and mass transfer in a biological tissue modeled as a sphere or cylinder, diffusion of ions through a cylindrical nerve fiber etc. In all of these cases, the physical domain is naturally radial, and the Cartesian model is just the incorrect geometric framework. Because the region through which the flux flows varies with radius, using a Cartesian model for a spherical or cylindrical domain will results in systematic mistakes in the flux. The geometrical correction in diffusion process can be handled appropriately by considering radial diffusion operator
\begin{align*}
    \Delta_{\nu}=\,\partial_{zz}+\frac{\nu}{z}\,\frac{\partial}{\partial z}=\,\frac{1}{z^{\nu}}\frac{\partial}{\partial z}\bigg(z^{\nu}\,\frac{\partial}{\partial z}\bigg).
\end{align*}
The term $\frac{\nu}{z}$ is a very crucial correction as it represents geometric spreading or convergence of flux with varying cross section of the medium. So studying the radial system is therefore more than merely a technical generalization. It yields physically realistic solutions and essentially distinct symmetry structures.
Motivated by this fact, we propose to study nonlinear heat and mass transfer with cross-diffusion effects over varying cross section geometry: 
\begin{equation}{\label{heq:1}}
\begin{aligned}
C_1(u,v)\,u_t = \dfrac{1}{z^{\nu}} 
\dfrac{\partial}{\partial z}\!\bigg[z^{\nu}\!\left(K(u,v)\,u_z + L(u,v)\,v_z\right)\bigg], \\
C_2(u,v)\,v_t = \dfrac{1}{z^{\nu}} 
\dfrac{\partial}{\partial z}\!\bigg[z^{\nu}\!\big(M(u,v)\,u_z + D(u,v)\,v_z\big)\bigg].
\end{aligned}
\end{equation}
The cartesian version of similar system is recently studied for Lie group classification \cite{stepanova2025lie}. This system covers Soret and Dufour effects and the authors has determined all admissible forms of the classified functions and obtained relationships between the classified parameters and the corresponding admitted generators.

\section{Lie group symmetries of the governing equation}
We investigate the heat conduction equation \eqref{heq:1} for Lie symmetries and it is equivalent to
\begin{equation}{\label{heq:2}}
\begin{aligned}
G_{1}
\equiv{}&
C_{1}(u,v)\,u_{t}
-K(u,v)\,u_{zz}
-L(u,v)\,v_{zz}
-K_{u}(u,v)\,u_{z}^{2}
-\bigl(K_{v}(u,v)+L_{u}(u,v)\bigr)\,u_{z}v_{z}\\
&-L_{v}(u,v)\,v_{z}^{2}
-\frac{\nu}{z}
\left(
K(u,v)\,u_{z}
+L(u,v)\,v_{z}
\right)
=0,
\\
G_{2}
\equiv{}&
C_{2}(u,v)\,v_{t}
-M(u,v)\,u_{zz}
-D(u,v)\,v_{zz}
-M_{u}(u,v)\,u_{z}^{2}
-\bigl(M_{v}(u,v)+D_{u}(u,v)\bigr)\,u_{z}v_{z}\\
&-D_{v}(u,v)\,v_{z}^{2}
-\frac{\nu}{z}
\left(
M(u,v)\,u_{z}
+D(u,v)\,v_{z}
\right)
=0.
\end{aligned}
\end{equation}
where $C_{1}(u,v),C_{2}(u,v)$ control the temporal response, storage, or inertial structure of the system $K(u,v),  D(u,v)$ represent self-diffusion coefficients. They describe the transport of each field due to its own gradient and 
$L(u,v), M(u,v)$ represent cross-diffusion coefficients. These terms describe transport induced by gradients of the other component.
are temperature dependent coefficients.

 We consider a one-parameter local Lie group of point transformations acting on
\((z,t,u,v)\), given by

\begin{align*}
\tilde z &= z+\varepsilon \xi(z,t,u,v)+O(\varepsilon^2),\quad
\tilde t = t+\varepsilon \tau(z,t,u,v)+O(\varepsilon^2),\\
\tilde u &= u+\varepsilon \eta(z,t,u,v)+O(\varepsilon^2),\quad
\tilde v = v+\varepsilon \phi(z,t,u,v)+O(\varepsilon^2),
\end{align*}
where \(\varepsilon\) is the group parameter, and $\xi, \tau, \eta, \phi
$
are the infinitesimals corresponding to \(z,t,u\), and \(v\), respectively.

The associated infinitesimal generator is
\begin{equation}
X=\xi(z,t,u,v)\frac{\partial}{\partial z}
+\tau(z,t,u,v)\frac{\partial}{\partial t}
+\eta(z,t,u,v)\frac{\partial}{\partial u}
+\phi(z,t,u,v)\frac{\partial}{\partial v}.
\end{equation}

For a system involving derivatives up to order \(n\), the prolonged generator is written as
\begin{equation}
\operatorname{pr}^{(n)}X
=
X
+
\sum_{\alpha=1}^{2}
\sum_{|J|=1}^{n}
\eta^{\alpha}_{J}
\frac{\partial}{\partial u^{\alpha}_{J}}, \quad (u^1,u^2)=(u,v),\,
(x^1,x^2)=(z,t)
\end{equation}
where $J$ denotes a multi-index corresponding to derivatives with respect to $z$ and $t$.

The prolonged coefficients are given by
\begin{equation*}
\eta^\alpha_J
=
D_J
\left(
Q^\alpha
\right)
+
\sum_{i=1}^{2}
\xi^i u^\alpha_{J,i},
\end{equation*}
where$(\xi^1,\xi^2)=(\xi,\tau),(Q^1,Q^2)=(Q^u,Q^v)$
with characteristics
\begin{align*}
Q^u = \eta-\xi u_z-\tau u_t,
Q^v = \phi-\xi v_z-\tau v_t.
\end{align*}

Here \(D_J\) denotes the total derivative operator corresponding to the multi-index \(J\). In particular,
\begin{align*}
D_z &= \frac{\partial}{\partial z}
+u_z\frac{\partial}{\partial u}
+v_z\frac{\partial}{\partial v}
+u_{zz}\frac{\partial}{\partial u_z}
+v_{zz}\frac{\partial}{\partial v_z}
+u_{zt}\frac{\partial}{\partial u_t}
+v_{zt}\frac{\partial}{\partial v_t}
+\cdots,\\
D_t &= \frac{\partial}{\partial t}
+u_t\frac{\partial}{\partial u}
+v_t\frac{\partial}{\partial v}
+u_{zt}\frac{\partial}{\partial u_z}
+v_{zt}\frac{\partial}{\partial v_z}
+u_{tt}\frac{\partial}{\partial u_t}
+v_{tt}\frac{\partial}{\partial v_t}
+\cdots.
\end{align*}

Suppose the given system of partial differential equations is written as
\begin{equation}
\Delta_\beta
\left(
z,t,u,v,u_z,u_t,v_z,v_t,\ldots
\right)=0,
\qquad
\beta=1,2.
\end{equation}

Then the invariance criterion is
\begin{equation}
\operatorname{pr}^{(n)}X(G_\beta)\bigg|_{G_\beta=0}=0,
\qquad
\beta=1,2.
\end{equation}

This condition yields the determining equations for the infinitesimals $\xi, \tau,\eta, \phi$, equating to zero the coefficient of derivative of $u$ and $v$ following determining equations are obtained
\begin{subequations}{\label{heq:3}}
\begin{align}
&\xi_{u}=\xi_{v}=0,\qquad
\tau_{z}=\tau_{u}=\tau_{v}=0,
\label{DE0}
\\[1ex]
&C_{1}\eta_{t}
-K\eta_{zz}
-L\phi_{zz}
-\frac{\nu}{z}K\eta_{z}
-\frac{\nu}{z}L\phi_{z}
=0,
\label{DE1}
\\[1ex]
&\frac{\eta C_{1u}K_{u}}{C_{1}}
+\frac{\phi C_{1v}K_{u}}{C_{1}}
-K\eta_{uu}
-\eta K_{uu}
-\tau_{t}K_{u}
-\eta_{u}K_{u}
-\phi K_{uv}
+2\xi_{z}K_{u}
=0,
\label{DE2}
\\[1ex]
&-\eta K_{uv}
-\eta L_{uu}
-L\tau_{u}
+2L\xi_{u}
-L_{u}\phi_{u}
-\phi K_{vv}
-\phi L_{uv}
-\phi_{v}K_{v}
\nonumber\\
&\qquad
+\frac{\eta C_{1u}L_{uu}}{C_{1}}
+\frac{\eta C_{1u}K_{v}}{C_{1}}
+\frac{L\phi C_{1v}}{C_{1}}
+\frac{\phi C_{1v}K_{v}}{C_{1}}
+L\eta_{uv}
+2\xi_{z}K_{v}
-\tau_{t}K_{v}
\nonumber\\
&\qquad
-2K\eta_{uv}
-\phi_{u}L_{v}
-2\eta_{v}K_{u}
=0,
\label{DE3}
\\[1ex]
&\frac{\eta C_{1u}L_{v}}{C_{1}}
+\frac{\phi C_{1v}L_{v}}{C_{1}}
-L\phi_{vv}
-\eta L_{uv}
-\tau_{t}L_{v}
-\phi L_{vv}
+2\xi_{z}L_{v}
\nonumber\\
&\qquad
-2L\phi_{v}
+\eta_{u}L_{v}
-\eta_{v}L_{u}
-\eta_{v}K_{v}
=0,
\label{DE4}
\\[1ex]
&\frac{K\eta C_{1u}}{C_{1}}
+\frac{K\phi C_{1v}}{C_{1}}
-\eta K_{u}
-\phi K_{v}
+2K\xi_{z}
-K\tau_{t}
-L\phi_{u}
=0,
\label{DE5}
\\[1ex]
&\frac{L\eta C_{1u}}{C_{1}}
+\frac{L\phi C_{1v}}{C_{1}}
-\phi L_{v}
-L\tau_{t}
+2L\xi_{z}
+L\eta_{u}
\nonumber\\
&\qquad
-L\phi_{v}
-\eta L_{u}
-K\eta_{v}
=0,
\label{DE6}
\\[1ex]
&\frac{L\eta C_{1u}\nu}{zC_{1}}
+\frac{L\phi C_{1v}\nu}{zC_{1}}
+\frac{L\eta_{v}\nu}{z}
+\frac{L\xi_{z}\nu}{z}
+\frac{L\xi\nu}{z^{2}}
\nonumber\\
&\qquad
-\frac{\phi L_{v}\nu}{z}
-\frac{L\tau_{t}\nu}{z}
-\frac{L\phi_{v}\nu}{z}
-\frac{\eta L_{u}\nu}{z}
-2L\phi_{vz}
-\phi_{z}L_{v}
\nonumber\\
&\qquad
-\eta_{z}L_{u}
-\eta_{z}K_{v}
+L\xi_{zz}
=0,
\label{DE7}
\\[1ex]
&\frac{K\eta C_{1u}\nu}{zC_{1}}
+\frac{K\phi C_{1v}\nu}{zC_{1}}
+\frac{K\xi_{z}\nu}{z}
+\frac{K\xi\nu}{z^{2}}
-\frac{K\tau_{t}\nu}{z}
\nonumber\\
&\qquad
-\frac{\eta K_{u}\nu}{z}
-\frac{\phi K_{v}\nu}{z}
-2K\eta_{uz}
-\phi_{z}L_{u}
-\phi_{z}K_{v}
-2\eta_{z}K_{u}
\nonumber\\
&\qquad
+K\xi_{zz}
-C_{1}\xi_{t}
=0,
\label{DE8}
\\[2ex]
&C_{2}\phi_{t}
-M\eta_{zz}
-D\phi_{zz}
-\frac{\nu}{z}M\eta_{z}
-\frac{\nu}{z}D\phi_{z}
=0,
\label{DE9}
\\[1ex]
&\tau_{t}M_{u}
-M\eta_{uu}
-\eta M_{uu}
-\eta_{u}M_{u}
-\phi M_{uv}
-2\eta_{u}M
\nonumber\\
&\qquad
+2\xi_{z}M_{u}
-\phi_{u}M
-\frac{C_{2u}M\eta}{C_{2}}
-\frac{\phi C_{2v}M_{u}}{C_{2}}
=0,
\label{DE10}
\\[1ex]
&\tau_{t}M_{v}
+\tau_{t}D_{u}
-M\eta_{uv}
+2M_{v}\xi_{z}
-M\phi_{v}
-D\eta_{uu}
\nonumber\\
&\qquad
+2D_{u}\xi_{z}
-2D\phi_{u}
-M_{uv}\eta
-M_{vv}\phi
-D_{uu}\eta
-D_{uv}\phi
\nonumber\\
&\qquad
-M_{u}\eta_{v}
-D_{u}\eta_{u}
-\phi M_{v}
-\phi D_{u}
-2N_{v}\phi_{u}
\nonumber\\
&\qquad
-\frac{C_{2u}N\eta}{C_{2}}
-\frac{\phi C_{2v}N_{u}}{C_{2}}
-\frac{M\phi C_{2v}}{C_{2}}
=0,
\label{DE11}
\\[1ex]
&\tau_{t}D_{v}
-\phi D_{v}
-D\phi_{vv}
-\eta D_{uv}
-\phi D_{vv}
-3\phi_{v}D_{v}
\nonumber\\
&\qquad
-\eta_{v}N_{u}
+2\xi_{z}D_{v}
-\frac{C_{2u}D\eta}{C_{2}}
-\frac{\phi C_{2v}D_{v}}{C_{2}}
=0,
\label{DE12}
\\[1ex]
&\tau_{t}M
-M\eta_{u}
+2\xi_{z}M
-M\phi_{v}
-M_{u}\eta
-M_{v}\phi
\nonumber\\
&\qquad
-\frac{C_{2u}M\eta}{C_{2}}
-\frac{\phi C_{2v}M}{C_{2}}
+\phi_{u}M
=0,
\label{DE13}
\\[1ex]
&\tau_{t}D
-N_{u}\eta
+2\xi_{z}D
-\phi D_{v}
-2\phi_{v}D
\nonumber\\
&\qquad
-\frac{C_{2u}D\eta}{C_{2}}
-\frac{\phi C_{2v}D}{C_{2}}
=0,
\label{DE14}
\\[1ex]
&\frac{D\eta C_{2u}\nu}{zC_{2}}
-\frac{D\phi C_{2v}\nu}{zC_{2}}
+\frac{\xi D\nu}{z^{2}}
+\frac{\tau_{t}D\nu}{z}
+\frac{\xi_{z}D\nu}{z}
\nonumber\\
&\qquad
-\frac{\eta D_{u}\nu}{z}
-\frac{\phi D_{v}\nu}{z}
-\frac{2\phi_{v}D\nu}{z}
-\eta_{z}N_{u}
-2D\phi_{vz}
\nonumber\\
&\qquad
-\phi_{z}D_{v}
+D\xi_{zz}
=0,
\label{DE15}
\\[1ex]
&\frac{M\eta C_{2u}\nu}{zC_{2}}
-\frac{\phi C_{2v}M\nu}{zC_{2}}
-\frac{\eta_{u}M\nu}{z}
+\frac{\xi M\nu}{z^{2}}
+\frac{\tau_{t}M\nu}{z}
+\frac{\xi_{z}M\nu}{z}
\nonumber\\
&\qquad
-\frac{\eta M_{u}\nu}{z}
-\frac{\phi M_{v}\nu}{z}
-\frac{\phi_{v}M\nu}{z}
-M_{u}\eta_{z}
-M_{v}\phi_{z}
\nonumber\\
&\qquad
-2M\eta_{uz}
-D\phi_{z}
+M\xi_{zz}
-C_{2}\xi_{t}
=0.
\label{DE16}
\end{align}
\end{subequations}
\section{Classification of constitutive functions}
For valid classification, the determining equations \eqref{heq:3} cannot be solved at once, we treat functional parameters $C_{1}(u,v),C_{2}(u,v), K(u,v),  D(u,v), L(u,v), M(u,v)$ as material laws, and we can classify them by observing how they transform under Lie point symmetries. For physically and geometrically relevant classification, we need to focus on capacity matrix and diffusion/conductivity matrix defined as follows:
\begin{equation}
{C}(u,v)
=
\begin{pmatrix}
C_{1}(u,v) & 0
\\
0 & C_{2}(u,v)
\end{pmatrix},
{A}(u,v)
=
\begin{pmatrix}
K(u,v) & L(u,v)
\\
M(u,v) & D(u,v)
\end{pmatrix}.
\end{equation}
 and since the structure of \eqref{heq:1} could be a radial/cylindrical/spherical cross-diffusion system with the geometry parameter $\nu$, the factor ${z^{-\nu}}{\partial z}(z^{\nu})$ will be strongly influence the classification process.

To proceed for classification we consider a one-parameter point transformation generated by
\begin{align*}
    X=\,\tau(t,z,u,v)\frac{\partial}{\partial t}+\xi(t,z,u,vv)\frac{\partial}{\partial z}+\eta(t,z,u,v)\frac{\partial}{\partial u}+\phi(t,z,u,v)\frac{\partial}{\partial v}
\end{align*}
determining equations \eqref{DE0} implies
\begin{align*}
    \tau=\,\tau(t),\quad \xi=\,\xi(t,z)
\end{align*}
This will be our first simplification, and 
\begin{align*}
    \eta=\,\eta(t,z,u,v),\quad \phi=\,\phi(t,z,u,v)
\end{align*}
will remain general at this stage of classification. For brevity, we define constitutive function
\begin{align*}
    F=\,F(u,v)\in\{C_{1},C_{2},K,D,L,M\}
\end{align*}
that depends on $(u,v)$ only, this will enforce any nontrivial symmetry involving $u$ and $v$ to act only on $(u,v)$ space. Consider $(u,v)$ space part of generator
\begin{align*}
    \mathcal{Q}=\, \eta\,\frac{\partial}{\partial u}+\phi\,\frac{\partial}{\partial v},
\end{align*}
if this symmetry is independent of $t,z$ then
\begin{align*}
    \mathcal{Q}=\, \eta(u,v)\,\frac{\partial}{\partial u}+\phi(u,v)\,\frac{\partial}{\partial v}
\end{align*}
This vector field will define the transformation in the constitutive function.
\begin{case}\normalfont
When the constitutive function $F(u,v)$ is arbitrary function of $(u,v)$, there is no nontrivial transformation in $(u,v)$ space that could preserve the functional dependence among constitutive functions except for the case $\eta=\phi=0$. This will substantially simplify the determining system \eqref{heq:3}, for example, \eqref{DE5} simplifies to 
\begin{align}
    \label{heq:4}2\,\xi_{z}-\tau_{t}=0, \quad K\neq0.
\end{align}
\eqref{DE6} also reduce to \eqref{heq:4} but with $L\neq0$. Since $\tau=\,\tau(t)$
and $\xi=\,\xi(t,z)$, so $\xi_{z}$ must be independent of $z$, so 
\begin{align}
    \xi_{z}=\,a(t)\implies \xi=\,a(t)\,z+b(t)
\end{align}
plugging into \eqref{heq:4} will yield:
\begin{align} \tau=\,2\,\int a(t)dt+c_{0},\quad \xi=\,a(t)\,z+b(t).
    \label{heq:5}
\end{align}
The rest of the equations are either consistent or identically zero, except for the equation \eqref{DE8} that will impose compatibility condition as follows:
\begin{align}
   \label{heq:6} \frac{K\nu b(t)}{z^{2}}-C_{1}a'(t)z-C_{1}b'(t)=0.
\end{align}
The geometric parameter $\nu$ can have different values, like $\nu=0$ for cartesian, $\nu=1$ for cylindrical, $\nu=2$ for spherical and $\nu\neq0$ for generalized taper. No new symmetry will be obtained for $\nu=1, 2$, so we divided the classification of geometric parameter into two sub-cases.
\begin{subcase}\normalfont
When $\nu\neq0.$ The coefficient of $z^{-2}$ in \eqref{heq:6} gives 
\begin{align*}
    K\nu b(t)=0 \implies  b(t)=0 \quad \text{for}\; K\neq0
\end{align*}
the coefficient of $z$ gives $-C_{1}a'(t)=0$, thus $a(t)=c_{1}$, we get 
\begin{align*}
    \tau=\,2c_{1}+c_{0},\quad \xi=\,c_{1}\,z
\end{align*}
and corresponding symmetries:
\begin{align*}
    X_{1}=\,\frac{\partial}{\partial t},\;X_{2}=\,2\,t\,\frac{\partial}{\partial t}+z\,\frac{\partial}{\partial z}
\end{align*}

\end{subcase}
\begin{subcase}\normalfont
    When $\nu=0.$ Then there will be no $z^{-2}$ term and 
    \begin{align*}
        -C_{1}a'(t)z-C_{1}b'(t)=0
    \end{align*}
\end{subcase}
must hold for all z, that is, we get
\begin{align*}
    a'(t)=0\quad b'(t)=0
\end{align*}
therefore from \eqref{heq:5}
\begin{align}
    \xi=\,c_{1}\,z+c_{2},\quad \tau=\, 2\,c_{1}\,t+c_{0}
\end{align}
\end{case}
and corresponding symmetries:
\begin{align*}
    X_{1}=\,\frac{\partial}{\partial t},\;X_{2}=\,\frac{\partial}{\partial z},\;X_{3}=\,2\,t\,\frac{\partial}{\partial t}+z\,\frac{\partial}{\partial z}
    \end{align*}

\begin{case}\normalfont
       When for any constitutive function $F=\,F(u,v)\in\{C_{1},C_{2},K,D,L,M\}$ 
       \begin{align*}
           F=F(s),\quad \text{for}\; s=\,a\,u+b\,v.
       \end{align*}
       Then $F$ will be invariant under $Q=\,b\partial_{u}-a\,\partial_{v}$, and this will give use freedom to choose:
       \begin{align*}
           \eta=\,c_{3}b,\quad \phi=\,-c_{3}a.
       \end{align*}
       The first simplification we get is:
       \begin{align*}
           &\eta_{t}=0,\eta_{z}=0,\eta_{u}=0,\eta_{v}=0\\
          & \phi_{t}=0,\phi_{z}=0,\phi_{u}=0,\phi_{v}=0.
       \end{align*}
       The determining equation \eqref{DE5} simplifies to following:
 \begin{align*}
     \frac{K}{C_{1}}\big(\eta \,C_{1u}+\phi\, C_{1v}\big)-\big(\eta \,K_{u}+\phi\, K_{v}\big)+K(2\,\xi_{z}-\tau_{t})=0.
 \end{align*}      
  The invariance condition $Q(C_{1})=Q(K)=0$ yields:
  \begin{align}
      \tau=\,2\,\int a(t)dt+c_{0},\quad \xi=\,a(t)\,z+b(t)
  \end{align}
  The remaining determining equations will meet the similar consequence and some will lead to compatibility conditions. 
  \begin{subcase}
      \normalfont When $\nu\neq0.$ Then compatibility condition will yield $b(t)=0,\, a'(t)=0$ and we get
      \begin{align}
    \xi=\,c_{1}\,z,\quad \tau=\, 2\,c_{1}\,t+c_{0},\quad \eta=\,c_{3}b,\quad \phi=\,-c_{3}a.
\end{align}
and corresponding symmetries:
\begin{align*}
    X_{1}=\,\frac{\partial}{\partial t},\;X_{2}=\,2\,t\,\frac{\partial}{\partial t}+z\,\frac{\partial}{\partial z},\; X_{3}=\,b\,\frac{\partial}{\partial u}-a\,\frac{\partial}{\partial v}
\end{align*}
  \end{subcase}
  
  \begin{subcase}
      \normalfont
      When $\nu=0$. Then $b(t)=\,c_{2},\,a(t)=c_{1}$, and resulting symmetries will be 
      \begin{align*}
    X_{1}=\,\frac{\partial}{\partial t},\;X_{2}=\,\frac{\partial}{\partial z},\;X_{3}=\,2\,t\,\frac{\partial}{\partial t}+z\,\frac{\partial}{\partial z},\; X_{4}=\,b\,\frac{\partial}{\partial u}-a\,\frac{\partial}{\partial v}
\end{align*}
  \end{subcase}
 \end{case}
 \begin{case}\normalfont
     When constitutive function $F=F(u)$, that is, it depends only on $u$. Then following symmetries will be obtained.
     \begin{align*}
    X_{1}=\,\frac{\partial}{\partial t},\;X_{2}=\,2\,t\,\frac{\partial}{\partial t}+z\,\frac{\partial}{\partial z},\; X_{3}=\,\frac{\partial}{\partial u},\quad \text{for}\,\nu\neq0
\end{align*}
and 
\begin{align*}
    X_{1}=\,\frac{\partial}{\partial t},\;X_{2}=\,\frac{\partial}{\partial z},\;X_{3}=\,2\,t\,\frac{\partial}{\partial t}+z\,\frac{\partial}{\partial z},\; X_{4}=\,\frac{\partial}{\partial u}, \quad \text{for}\,\nu=0
    \end{align*}
    Similarly,  the case when $F=F(v)$ can also be discussed.
 \end{case}
 \begin{case}\normalfont
 When constitutive function $F(u,v)$ is homogeneous in $u$ and $v$, then the natural state-space part of the symmetry generator is a scaling in the dependent variables and 
 \begin{align}
    \label{heq:8} p\,u\,\frac{\partial F}{\partial u}+q\,v\, \frac{\partial F}{\partial v}=\,\lambda_{F}\,F.
 \end{align}
    Then differential operator for scaling in dependent variables will be:
    \begin{align}
        Q=\,p\,u\,\frac{\partial }{\partial u}+q\,v\, \frac{\partial }{\partial v}
    \end{align}
    so that for constitutive function to be compatible with scaling the weighted homogeneity condition \eqref{heq:8}  will implies $Q(F)=\,\lambda_{F}\,F$. The general solution for \eqref{heq:8} is 
    \begin{align*}
        F(u,v)=\,u^{\frac{\lambda_{F}}{p}}\mathcal{F}\left(v\,u^{\frac{q}{p}}\right),\;p\neq0.
    \end{align*}
    For simple power law solution $F=\,F_{0}\,u^{\alpha_{F}}\,v^{\beta_{F}}$, so that $\lambda_{F}=\,p\,\alpha_{F}+q\,\lambda_{F}$. The initial simplification we have here is 
    \begin{align*}
        \eta=\,p\,u,\;\phi=\,q\,v
    \end{align*}
    putting this into determining equation \eqref{DE5} provides:
    \begin{align*}
        \frac{K}{C_{1}}\,Q(C_{1})-Q(K)+K(2\,\xi_{z}-\tau_{t})=0
    \end{align*}
    for $Q(C_{1})=\,\lambda_{C_{1}}\,C_{1}, Q(K)=\,\lambda_{K}\,K$ this further simplifies to 
    \begin{align}
        \label{heq:7}\lambda_{C_{1}}-\lambda_{K}=\tau_{t}-2\,\xi_{z},  
    \end{align}
    and similarly the determining equation \eqref{DE6} reduce to
    \begin{align}
        \label{heq:9}\tau_{t}-2\,\xi_{z}=\lambda_{C_{1}}-\lambda_{L}+p-q
    \end{align}
    To ensure consistency among \eqref{heq:7} and \eqref{heq:9} we must have:
    \begin{align}
        \label{heq:10}\lambda_{K}=\lambda_{L}-p+q.
    \end{align}
    This formula explains the scaling of cross diffusion coefficient $L$ relative to $K$, and from remaining equation we get:
    \begin{equation}{\label{heq:11}}
    \begin{aligned}
        \tau_{t}-2\,\xi_{z}=\lambda_{C_{2}}-\lambda_{D},\\
        \lambda_{M}=\lambda_{D}+q-p.
    \end{aligned}
    \end{equation}
     From equations \eqref{heq:9}, \eqref{heq:10} and \eqref{heq:11}, the classifying equations can be obtained as follows:
     \begin{align*}
         \lambda_{K}=\lambda_{C_{1}}-\sigma,\; \lambda_{L}=\lambda_{C_{1}}+p-q-\sigma,\\
         \lambda_{D}=\lambda_{C_{2}}-\sigma,\;\lambda_{M}=\lambda_{C_{2}}+q-p-\sigma,
     \end{align*}
     where notation $\sigma=\tau_{t}-2\,\xi_{z}$ is used and is constant. Since $\xi_{z}$ depends only on $t$, so we get
     \begin{align}
         \label{heq:12}\xi=a(t)\,z+b(t),\;\tau=\int (2\,a(t)+\sigma)dt+c_{0}.
     \end{align}
     The splitting of geometry will yield two different types symmetries  and calculation are same as done before, we have:
     \begin{align*}
         \nu\neq0 \implies \begin{cases}
         X_{1}=\frac{\partial}{\partial t}\\
         X_{2}=2\,t\,\frac{\partial}{\partial t}+z\,\frac{\partial}{\partial z}\\
         X_{3}=\sigma\,t\,\frac{\partial}{\partial t}+p\,u\frac{\partial}{\partial u}+q\,v\frac{\partial}{\partial v}
     \end{cases}
     \end{align*}
     and for $\nu=0$ additional symmetry $\partial_{z}$ will be obtained.
 \end{case}
 \begin{case}\normalfont
 When constitutive function $F(u,v)$ is exponential function of $u$ and $v$, then obvious operator will be
 \begin{align*}
     Q=\,p\,\frac{\partial}{\partial u}+q\,\frac{\partial}{\partial v}
 \end{align*}
     that will generate translation  $u\mapsto u+p\,\epsilon$ and $v\mapsto v+q\,\epsilon$, such that 
     \begin{align}
         \label{heq:13}Q(F)=\,\lambda_{F}\,F.
     \end{align}
     For $F(u,v)=\,F_{0}\,\text{exp}(\alpha_{F}\,u+\beta_{F}\,v)$, then \eqref{heq:13} will give $\lambda_{F}=\,p\,\alpha_{F}+q\,\beta_{F}$. The initial simplification in this case is 
     \begin{align*}
         \eta=\,p,\;\phi=\,q.
     \end{align*}
     The determining equations \eqref{DE5} give:
     \begin{align}
         \label{heq:14}\tau_{t}-2\,\xi_{z}=\,\lambda_{C_{1}}-\lambda_{K},
     \end{align}
     and remaining determining equation will give:
     \begin{align}
         \label{heq:15}\sigma=\,\lambda_{C_{1}}-\lambda_{L}=\lambda_{C_{2}}-\lambda_{M}=\lambda_{C_{2}}-\lambda_{D}
     \end{align}
     The splitting of geometry will yield two different types symmetries  and calculation are same as done before, we have:
     \begin{align*}
         \nu\neq0 \implies \begin{cases}
         X_{1}=\frac{\partial}{\partial t}\\
         X_{2}=2\,t\,\frac{\partial}{\partial t}+z\,\frac{\partial}{\partial z}\\
         X_{3}=\sigma\,t\,\frac{\partial}{\partial t}+p\,\frac{\partial}{\partial u}+q\,\frac{\partial}{\partial v}
     \end{cases}
     \end{align*}
     and for $\nu=0$ additional symmetry $\partial_{z}$ will be obtained.
 \end{case}
 \begin{case}\normalfont
     When constitutive function $F(u,v)$ is constant, then splitting of geometry will give following symmetries:
     \begin{align*}
         \nu\neq0 \implies \begin{cases}
             X_{1}=\,\frac{\partial}{\partial t}\\
             X_{2}=\,2\,t\,\frac{\partial}{\partial t}+z\,\frac{\partial}{\partial z}
         \end{cases}
     \end{align*}
     and 
     \begin{align*}
         \nu=0\implies\begin{cases}
             X_{1}=\,\frac{\partial}{\partial t},\\
             X_{2}=\,\frac{\partial}{\partial z},\\
             X_{3}=\,2\,t\,\frac{\partial}{\partial t}+z\,\frac{\partial}{\partial z},\\
             X_{4}=\,2\,t\,\frac{\partial}{\partial z}-\frac{z}{\kappa}\left(u\,\frac{\partial}{\partial u}+v\,\frac{\partial}{\partial v}\right),\;\text{Galilean type symmetry},\\
             X_{5}=\,4\,t^{2}\frac{\partial}{\partial t}+4\,tz\,\frac{\partial}{\partial z}-\left(\frac{z^{2}+2\kappa t}{\kappa}\right)\left(u\,\frac{\partial}{\partial u}+v\,\frac{\partial}{\partial v}\right),\;\text{projective symmetry}
         \end{cases}
     \end{align*}
 \end{case}
 \begin{case}\normalfont
 When the system is diagonally uncoupled, that is, for $L=M=0$. For power law diffusivity we may assume:
 \begin{align*}
     &C_{1}=\,1,\;K=\,u^{m}\\
     &C_{2}=\,1,\; D=\,v^{n}.
 \end{align*}
  We can try symmetry
  \begin{align}
      X=\,\alpha\,t\,\frac{\partial}{\partial t}+\beta\,z\,\frac{\partial}{\partial z}+\gamma\,u\,\frac{\partial}{\partial u}+\delta\,v\frac{\partial }{\partial v}
  \end{align}
  which is equivalent to scaling \[
t \to \lambda^{\alpha} t,\qquad
z \to \lambda^{\beta} z,\qquad
u \to \lambda^{\gamma} u,\qquad
v \to \lambda^{\delta} v.
\]
Applying this scaling in both equations \eqref{heq:1} will give 
$$\alpha=\,-mn,\;\beta=\,0,\;\gamma=\,n,\;\delta=\,m.$$
So we get extra symmetry $$X=\, -mn\,t\,\frac{\partial}{\partial t}+n\,u\,\frac{\partial}{\partial u}+m\,v\,\frac{\partial}{\partial v}$$
along with kernel symmetry 
$$X=\,2\,t\,\frac{\partial}{\partial t}+z\,\frac{\partial}{\partial z}$$
and for $\nu=0$, the additional symmetry $\partial z$ can also be included.
 \end{case}
 \begin{case}\normalfont
     Reciprocal cross diffusion $L=M$. This case is important because it is physically relevant, but it won't give extra symmetries unless the constitutive function $F$ is invariant under some $Q$, otherwise it will only restrict the functional class. We can try rotational symmetry:
     \begin{align}
         Q=\,v\,\frac{\partial}{\partial u}-v\,\frac{\partial}{\partial v}
     \end{align}
     that generates transformation $u\mapsto u\,\cos\epsilon+v\,\sin\epsilon, v\mapsto -u\,\sin\epsilon+v\,\cos\epsilon$, a clear invariant is $r^{2}=u^{2}+v^{2}$, so that for $F=F(r^2)$ the invariance condition holds as $Q(F)=0$. But the mere dependence of constitutive functions on $r$ is not sufficient,  the diffusion matrix ${A}$ should also transforms covariantly under rotations, that is we must have 
     \begin{align}
         {A}(\mathbf{u})=\,a(r)I+b(r)\mathbf{u}\mathbf{u}^{T}=\,a(r)I+b(r)\, \begin{pmatrix}
u^{2} & uv
\\
uv & v^{2}
\end{pmatrix}
     \end{align}
     where $a(r)I$ gives isotropic diffusion equally along all directions and $b(r)\mathbf{u}\mathbf{u}^{T}$ modifies diffusion along radial directions. For rotation $\mathbf{u}\mapsto R\mathbf{u}$, the diffusion matrix  transforms covariantly as ${A}(R\mathbf{u})=R({A(\mathbf{u})})R^{T}$. So initial simplification at this stage is 
     \begin{align*}
         \eta=\,v,\;\phi=\,-u
     \end{align*}
     solution of  the determining equations \eqref{heq:3} under this simplification will provide following symmetries under geometrical splitting:
     \begin{align*}
         \nu\neq0 \implies \begin{cases}
         X_{1}=\frac{\partial}{\partial t}\\
         X_{2}=2\,t\,\frac{\partial}{\partial t}+z\,\frac{\partial}{\partial z}\\
         X_{3}=v\,\frac{\partial}{\partial u}-u\,\frac{\partial}{\partial v}
     \end{cases}
     \end{align*}
     and for $\nu=0$ additional symmetry $\partial_{z}$ will be obtained.
 \end{case}
 The symmetries obtained so far are exactly consistent with geometry of governing equations \eqref{heq:1}, for example, the system is autonomous in time so time translation 
 \begin{align*}
     X_{1}=\, \frac{\partial}{\partial t}
 \end{align*}
 was natural symmetry. The diffusion operator of the type 
 \begin{align*}
     \frac{\partial^{2}}{\partial z^{2}}+\frac{\nu}{z}\,\frac{\partial}{\partial z}
 \end{align*}
 has parabolic  scaling structure: terms with first order time derivatives scale like one power of time and second order diffusion terms like $u_{zz}, v_{zz}$ and $z^{-1}u_{z}, z^{-1}v_{z}$ scale like two power of space which is equivalent to parabolic dilation 
 \begin{align*}
     X_{2}=\,2\,t\,\frac{\partial}{\partial t}+z\,\frac{\partial}{\partial z}
 \end{align*}
 but spatial translation will cease to exist for $\nu\neq0$ because of the factor $\frac{1}{z}$, it will exist only when $\nu=0$, therefore
 \begin{align*}
     X_{3}=\, \frac{\partial}{\partial z}
 \end{align*} was admitted for the cartesian case. The possible extra symmetries arising from the certain forms of the constitutive function $F(u,v)$ are also geometrically natural. When constitutive function  depends only on the invariant $s=\,a\,u+b\,v$, then the operator 
 \begin{align*}
     X=\,b\,\frac{\partial}{\partial u}-a\,\frac{\partial}{\partial v}
 \end{align*}
 will leave $F(u,v)=\,a\,u+b\,v$ invariant, so is also a natural and geometrically consistent symmetry. If the constitutive functions are homogeneous, the equations may admit a dependent-variable scaling symmetry, because homogeneous functions transform multiplicatively under rescaling of $(u,v)$. Similarly, the rotational symmetry 
 \begin{align*}
     X=\,v\frac{\partial}{\partial u}-u\,\frac{\partial}{\partial v}
 \end{align*}
 is also admitted because constitutive functions are invariant under rotation in $(u,v)$-plane. In addition to these natural symmetries, we have also tried to find some extra symmetries by enforcing affine transformations in $\eta$ and $\phi$ as follows:
 \begin{align*}
     \eta=&\,a_{11}\,u+a_{12}\,v+b_{1}+\psi_{1}(z,t),\\
     \phi=&\,a_{21}\,u+a_{22}\,v+b_{2}+\psi_{2}(z,t).
 \end{align*}
 But the nonlinear differential dependence of constitutive functions on $(u,v)$ forced nonhomogeneous terms $\psi_{1}(z,t),\psi_{1}(z,t)$ to vanish. The nonlinear dependence of the constitutive functions on $u$ and $v$ strongly restricts the possibility of additional symmetries. The determining equations \eqref{heq:3} are heavily constrained, which makes even the most possible transformations of $u$ and $v$ ruled out. Only those transformations survived that leave constitutive functions invariant or transform them in a simpler way. We can expect larger Lie algebra only in case of constant-coefficient linear systems, diagonal uncoupled systems reducible to scalar nonlinear diffusion equations where system essentially behaves like single diffusion equation or  in case of rank-deficient diffusion matrices where determinant of diffusion matrix vanishes or when system can be transformed to vector heat equation. However, these scenarios require additional restrictive assumptions that will destroy the overall coupled nonlinear structure and that would not represent typical characteristics of the current nonlinear functional model.
\section{Conclusion}
In this paper, a complete Lie point symmetry classification has been carried out for a coupled nonlinear cross-diffusion system with a varying cross-section geometry controlled by the parameter $\nu$. The system involves two capacity functions $C_{1}(u,v), C_{2}(u,v)$ and four constitutive function $K(u,v), L(u,v), M(u,v), D(u,v)$ which depend nonlinearly on dependent variables $u$ and $v$. Applying the Lie invariance criterion to the  system \eqref{heq:1} resulted in sixteen determining equations for the infinitesimal generators $\xi, \tau, \eta, \phi$. Proceeding in systematic manner it is proved that, universally $\tau$ depends only on $t$  and $\xi$ depends on $(z,t)$. The strong nonlinearity in remaining equations is handled with the operator $Q=\,\eta\,\partial_{u}+v\,\partial_{v}$ which act in state space. 

For arbitrary constitutive functions, the system admits natural symmetries such as $\partial_{t}$ and $2t\,\partial_{t}+z\,\partial_{z}$ which are equivalent to autonomous nature of system and natural balance between the time derivative and the second-order radial diffusion operator. For $\nu=0$, the geometric term $\frac{\nu}{z}$ get annihilated and  the translational invariance in $z$ is obtained, so $\partial_{z}$ is admitted. For $\nu\neq0$, the term $\frac{\nu}{z}$ destroys the spatial invariance, and this demonstrates the consequence of geometric  difference between cartesian and radial cases.

Extra symmetry $b\,\partial_{u}-a\,\partial_{v}$ is obtained when the constitutive function $F(u,v)$ depends on the invariant $a\,u+b\,v$. Furthermore, when $F(u,v)$ is homogeneous or exponential in type, then symmetries are obtained which are geometrically natural and directly related to the invariant structure of the constitutive functions.
In search of more symmetries,  general affine ansatz for the state-space part of the generator is also considered by assuming $\eta$ and $\phi$ to affine in $u$ and $v$ with addition of $\psi_{1}(z,t)$ and $\psi_{2}(z,t)$. But the strong nonlinear dependence of constitutive functions in determining equations forced the annihilation of  $\psi_{1}(z,t)$ and $\psi_{2}(z,t)$.

To summarise, the current system's Lie symmetry algebra is essentially maximal in the class of point symmetries for nonlinear coupled cross-diffusion with radial geometry. The resulting symmetries give a platform for creating symmetry reductions, invariant solutions, and conservation laws, which can be developed in future study.


 \bibliography{My.Bibtex.Library} 
 \bibliographystyle{elsarticle-num}

\end{document}